\documentclass[12pt]{iopart}

\usepackage{graphicx,epsfig,epstopdf}
\usepackage{graphicx}
\usepackage[center]{caption}

\begin{document}

\title[Quarkonium spectroscopy of the linear plus modified  Yukawa potential  ]{Quarkonium spectroscopy of the linear plus modified  Yukawa potential}

\author{Kaushal R Purohit $^{1*}$, Pooja Jakhad$^1$ and  Ajay Kumar Rai$^1$ }

\address{$^{1*}$ Department of Physics, Sardar Vallabhbhai National Institute of Technology, Surat, Gujarat-395 007, India. \\
}
\ead{kaushalsep1996@gmail.com}
\vspace{10pt}

\begin{abstract}

In this article, the linear plus modified Yukawa potential (LIMYP) is used as the quark-antiquark interaction potential for the approximate analytical bound state solution of the Klein–Gordon equation in three-dimensional space. The energy eigenvalues and associated wavefunction are obtained by solving the Klein–Gordon equation analytically using the Nikiforov–Uvarov (NU) method. The mass spectra of heavy mesons such as charmonium $(c\bar{c})$, bottomonium $(b\bar{b})$, and $b\bar{c}$ for various quantum states are obtained using the energy spectra expression. In comparison to experimental data, graphical modification of acquired mass spectra of heavy mesons with the parameter employed in the energy equation and the current potential provides good results.

\end{abstract}

%

\section{Introduction}
 Several researchers have used various approaches to get exact/approximate solutions of the Schrodinger, Dirac and Klein–Gordon equations using certain common potentials \cite{gc.1,gc.2,gc.3,gc.4,gc.5}. These methods are Nikiforov–Uvarov (NU) method \cite{gc.6,gc.6a},  asymptotic iteration method   (AIM) \cite{gc.7}, supersymmetric quantum mechanics (SUSYQM) \cite{gc.8,gc.9}, factorization method \cite{gc.10,gc.11}, the exact quantization rule method
\cite{gc.12} and Qiang–Dong proper quantization rule \cite{gc.13}, etc.

There has recently been a lot of interest in merging two or more potentials in the relativistic and non-relativistic regimes. A wider range of applications can be achieved by combining two or more physical potential models \cite{gc.14}.
We shall describe the behavior of quarkonium systems in this paper by selecting two different phenomenological potentials (liner plus modified Yukawa potential) and fitting the spectra with coefficient potentials.

Linear plus modified Yukawa potential (LIMYP)takes as

\begin{equation}  V(r)=A_{1} r+ \frac{A_{2} e^{-\alpha r}}{r}-\frac{A_{3} e^{-2 \alpha r}}{r^2}+A_{4} \label{eq:gc30}  \end{equation}

Where  potential strength such as $A_{1}, A_{2}, A_{3}, A_{4}$ and $A_{5}$ screening parameter as a $\alpha$

\begin{equation}e^{-\alpha r}=1-\frac{\alpha r}{1!}+\frac{\alpha^2 r^2}{2!}-\frac{\alpha^3 r^3}{3!} \label{eq:gc30a}\end{equation}
\begin{equation} e^{-2 \alpha r }= 1-\frac{2 \alpha r}{1!}+\frac{4 \alpha^2 r^2}{2!}-\frac{8 \alpha^3 r^3}{3!} \label{eq:gc30b} \end{equation}

In order to describe the potential to interact in the quark-antiquark system, we perform series expansions of the exponential term in Eq.(\ref{eq:gc30a}) and (\ref{eq:gc30b}) up to order three and substitute the results into Eq.(\ref{eq:gc30})

\begin{equation} V(r)=\frac{\eta_{1}}{r^2} + \frac{\eta_{2}}{r}+\eta_{3}r + \eta_{4}r^2+\eta_{5} \label{eq:gc3}  \end{equation}

Where
\[ \eta_{1}= -A_{3}  ,\eta_{2}=A_{2} + 2 A_{3} \alpha, \eta_{3}= A_{1}+\frac{A_{2}\alpha^2}{2}+\frac{4}{3}A_{3}\alpha^3 \]
\begin{equation} \eta_{4}= -\frac{A_{2} \alpha^3 }{6}, \eta_{5}=-A_{2} \alpha - 2 A_{3} \alpha^2+A_{4}\label{eq:gc3a}\end{equation}

In equation (\ref{eq:gc3}), the $2^{nd}$ term is coulomb potential that describes the short distance between quarks and the third term is a linear term for confinement feature.

Cornell potential is one of the potentials that has attracted a lot of interest in particle physics. It has been successfully employed in models describing heavy quark binding states \cite{gc.14a,gc.14b,gc.14c}. Schrodinger's solutions have received a lot of attention.
The SE and K-G equation has sparked researchers' curiosity when viewed through the framework of quarkonium interaction potentials such as the Cornell or Killingbeck potentials \cite{gc.15,gc.16,gc.17,gc.18}. Ibekwe et al. \cite{gc.19} used an exponential, generalized, anharmonic Cornell to solve the radial Schrödinger equation analytically and obtain the mass spectral formula for heavy quarkonium systems. Using the N–U method in the non-relativistic quark model Abu-Shady et al. \cite{gc.20} examined the masses and thermodynamic properties of heavy mesons. The WKB approximation method was used by Omugbe et al. \cite{gc.21} to derive the mass spectrum of mesons. Ciftci and Kisoglu \cite{gc.22} established energy eigenvalues for an exact SE and computed the mass of a heavy quark–antiquark system (quarkonium) using the asymptotic iteration method (AIM). E.omugbe et. al \cite{omugbe} studied Any $\ell$-State Energy of the Spinless Salpeter Equation Under the Cornell Potential by the WKB Approximation Method An Application to Mass Spectra of Mesons. N. R. Soni et al. also obtained  $Q\bar{Q}(Q \epsilon \{b, c\})$ spectroscopy using Cornell potential \cite{soni}. Theoretically, quarkonium physics is becoming increasingly significant as a result of the numerous experimental states that are currently available \cite{gc.23,gc.24,gc.25,gc.26,gc.27, pandya, Rai, Kher}. The recent searches for quarkonium in experiments worldwide such as BESIII, LHCb, CMS has been a driving force towards the present theoretical study \cite{CMS, BESIII, LHCb, JPAC}.

The following is how we structured our work: The theory of the generalized NU method is reviewed in section two. In Section 3, we calculate the energy eigenvalues and normalize the LIMYP wave function. The mass spectra of LIMYP are obtained in section 4. Results and discussion are presented in section 5. In Section 6, there is a brief conclusion.

\section{ Nikiforov-Uvarov (NU) method }
 Nikiforov and Uvarov \cite{gc.6} introduced the NU method, which uses a coordinate transformation of the type  $z=z(r)$, to convert a schrodinger  problem into a $2^{nd}$ order $ diff^{n} $ equation \cite{gc.10, gc.27a} .
 \begin{equation}  \psi^{''}(z)+\frac{\tilde{\tau} (z)}{\sigma (z)} \psi^{'}(z)+\frac{\tilde{\sigma}(z)}{\sigma^{2}(z)}\psi(z)=0 \label{eq:gc1}  \end{equation}
 Where $\tilde{\sigma}(z)$ and $\sigma(z)$ are polynomial of the $2^{nd}$ degree at most, and $\tilde{z}$ is first-degree polynomial. The transformation can be used to derive the exact solution of  Eq.(\ref{eq:gc1}).
  \begin{equation}  \psi(z)=\phi(z)y(z) \label{eq:gc1h}  \end{equation}
 This transformation converts Eq.(\ref{eq:gc1}) into a hypergeometric type equation of the form Eq.(6)
 \begin{equation}  \sigma(z)y^{''}(z)+\tau(z)y{'}(z)+\lambda y(z)=0 \end{equation}

  The logarithm derivative can be defined as the function $\phi(z)$.
 \begin{equation} \frac{\phi^{'}(z)}{\phi (z)}=\frac{\pi(z)}{\sigma (z)} \label{eq:gc1d} \end{equation}

  $\pi(z)$ is a first-degree polynomial at most. The hyper geometric function with polynomial solution given by Rodrigues relation as the second part $\psi(z)$ begins y(z) in Eqs.(7).
  \begin{equation}  y(z)=\frac{B_{nl}}{\rho(z)}\frac{d^{n}}{dz^{n}}\left[\sigma^{n}(z)\rho(z)\right] \label{eq:gc1e}\end{equation}
  Where $B_{nl}$ is the normalization constant and $\rho(z) $ is the weight function that meets the following criteria;
   \begin{equation} \left(\sigma(z)\rho(z)\rho(z)\right)^{'}=\tau(z)\rho(z) \label{eq:gc1f}\end{equation}
  Where also
  \begin{equation}  \tau(z)=\tilde{\tau}(z)+2\pi(z)\label{eq:gc1b} \end{equation}

  It was necessary for bound state solutions to have
 \begin{equation}  \tau^{'}(z)<0\end{equation}
   The following function $\pi(z)$ and parameter $\lambda$, respectively, can be used to generate the eigenfunction and eigenvalues:
  \begin{equation}  \pi(z)= \frac{\sigma^{'}(z)-\tilde{\tau}(z)}{2}\pm\sqrt{\left(\frac{\sigma^{'}(z)-\tilde{\tau}(z)}{2}\right)^2- \tilde{\sigma}(z)+k\sigma(z)} \label{eq:gc2}  \end{equation}

   and

   \begin{equation}  \lambda=k_{-}+\pi_{-}^{'}(z) \label{eq:gc2c} \end{equation}

   Setting the discriminant in the square root Eq. (\ref{eq:gc2}) equal to zero yields the value of $k$. As a result, the revised eigenvalues equation is as follows:

   \begin{equation}  \lambda+n\tau^{'}(z)+\frac{n(n-1)}{2}\sigma^{''}(z)=0, (n=0,1,2,....) \label{eq:gc2d} \end{equation}

\section{ Bound state solution of the Klein-Gordon equation with linear plus modified Yukawa potential (LIMYP)}
 For $\hbar=c=1$ in N-dimensions, the Klein-Gordon equation for a spin less particle is \cite{gc.28}
 \unboldmath   \[ \left[-\nabla^{2}+\left(M+S(r)\right)^2+\frac{(N+2\ell-1)(N+2\ell-3)}{4 r^2}\right]\psi(r,\theta,\varphi)=\]\begin{equation}[E_{n\ell}-V(r)]^2 \psi(r,\theta,\varphi)\end{equation}

  Where $\nabla^2$ represents Laplacian, $M$ represents reduced mass, $E_{n\ell}$ represents the energy spectrum, and $n$ and $\ell$ represent radial and orbital angular momentum quantum numbers, respectively. It is generally known that the wave function can be rewritten to satisfy the boundary condition.

  \begin{equation}\psi(r,\theta,\varphi)=\frac{R_{n\ell}}{r}Y_{\ell m}(\theta, \varphi)\end{equation}
 As seen below, the wave function's angular component might be separated, leaving only the radial part.

\begin{eqnarray}
\label{eq:gc4}
\nonumber 
  & \frac{d^2 R(r)}{dr^2} & +[(E_{n \ell}^2-M^2)+V^2(r)-S^2(r)-2(E_{n \ell}V(r)+M S(r)) \\
  &-&\frac{(N+2 \ell-1)(N+2 \ell-3)}{4r^2}] R(r)=0
\end{eqnarray}
 As a result, Eq.(\ref{eq:gc4}) becomes $V(r)=S(r)=2V(r)$ when the vector and scalar potentials are identical.

    \begin{eqnarray}
     \nonumber 
     &\frac{d^2 R(r)}{dr^2}&+\left[(E_{n \ell}^2-M^2)-2V(r)(E_{n \ell} + M)-\frac{(N+2 \ell-1)(N+2 \ell-3)}{4r^2}\right]\\
      &R(r)&=0 \label{eq:gc5}
      \end{eqnarray}

    Upon substituting Eq.(\ref{eq:gc3}) into Eq.(\ref{eq:gc5}), we obtain

\begin{eqnarray}
 \nonumber
& \frac{d^2 R(r)}{dr^2}+  \left[(E_{n \ell}-M^2)+(-\frac{ 2\eta_{1}}{r^2} + \frac{2 \eta_{2}}{r}- 2 \eta_{3}r+ 2\eta_{4}r^2- 2\eta_{5})(E_{n \ell} +M)\right] R(r) &\\
&-\left[\frac{(N+2 \ell-1)(N+2 \ell-3)}{4 r^2}\right]R(r)=0 & \label{eq:gc6}  \end{eqnarray}

We set $r$ equal $z$ in Eq. (\ref{eq:gc6}) to change the coordinate from $r$ to $z$.

\begin{equation} z=\frac{1}{r} \label{eq:gc7} \end{equation}

  This means that in  Eq.(\ref{eq:gc7}) he $2^{nd}$ derivative becomes;
  \begin{equation}  \frac{d^2 R(r)}{dr^2}= 2z^3 \frac{dR(z)}{dz}+z^4\frac{d^2 R(z)}{dz^2} \label{eq:gc8} \end{equation}

  Substituting Eq.(\ref{eq:gc7}) and (\ref{eq:gc8}) into Eq. (\ref{eq:gc6})

 \[\frac{d^2 R(z)}{dz^2}+\frac{2}{z}\frac{dR}{dz}+\frac{1}{z^4}\]
 \[\left[(E_{n \ell}^2-M^2)+\left(-2 \eta_{4} r^2+2 \eta_{3}r- \frac{2 \eta_{2}}{r}+\frac{2 \eta_{1}}{r^2}-2 \eta_{5}\right)(E_{n \ell}+M)\right]R(z)\]
 \begin{equation}\left[-\frac{(N+2 \ell-1)(N+2 \ell-3)z^2}{4}\right]R(z)=0 \label{eq:gc8c} \end{equation}

  Then, on the term, we recommend the following approximation scheme: $\frac{\eta_{1}^{'}}{x^{2}} $ and  $\frac{\eta_{2}^{'}}{x} $.

   Assume that the meson has a radius of $r_{0}$. The concept is therefore focused on the expansion of $\frac{\eta_{1}^{'}}{z^{2}} $ and  $\frac{\eta_{2}^{'}}{z} $ in power series around $r_{0}$; i.e., around $\delta\equiv\frac{1}{r_{0}}$, up to second order in X-space. This is comparable to the Pekeris approximation, which helps in the deformation of the centrifugal term such that the potential may be solved using the NU method \cite{gc.28}.

   It can be converted into a series of powers by setting $y=z-\delta$ and $y=0$;

  \begin{equation}\frac{\eta_{2}}{z}=\frac{\eta_{2}}{y+\delta}=\frac{\eta_{2}}{\delta\left(1+\frac{y}{\delta}\right)}=\frac{\eta_{2}}{\delta}\left(1+\frac{y}{\delta}\right)^{-1}\end{equation}

    which yields

 \begin{equation}  \frac{\eta_{2}}{z}= \eta_{2} \left(\frac{3}{\delta}-\frac{3 z}{\delta^2}+\frac{z^2}{\delta^2}\right)\label{eq:gc8a} \end{equation}

  similarly,

\begin{equation}\frac{\eta_{1}}{z^{2}}= \eta_{1}\left(\frac{6}{\delta^2}-\frac{8z}{\delta^3}+\frac{3 z^2}{\delta^4}\right) \label{eq:gc8b} \end{equation}

We get Eq. (\ref{eq:gc8c}) by substituting Eqs.(\ref{eq:gc8a}) and (\ref{eq:gc8b})

 \begin{equation} \frac{d^2 R(z)}{dz^2}+\frac{2z}{z^2}\frac{dR(z)}{dz}+\frac{1}{4}\left[-\varepsilon+\eta z-\gamma z^2 \right]R(z)=0 \label{eq:gc9} \end{equation}

Where

\begin{equation} -\varepsilon = (E_{n \ell}^2-M^2)-\frac{6 \eta_{3}}{\delta}(E_{n \ell}+M)+\frac{12 \eta_{4}}{\delta^2}(E_{n \ell}+M)-2 \eta_{5}(E_{n \ell}+M) \end{equation}
\begin{equation} \eta= 2 \eta_{3}(E_{n \ell}+M)+\frac{6 \eta_{3}}{\delta^2}(E_{n \ell}+M)-\frac{16 \eta_{4}}{\delta^3}(E_{n \ell}+M)\end{equation}
\begin{equation}\gamma= 2 \eta_{1}(E_{n \ell}+M)+\frac{2 \eta_{3}}{\delta^3}(E_{n \ell}+M)-\frac{6 \eta_{4}}{\delta^4}(E_{n \ell}+M)+\frac{(N+2 \ell-1)(N+2 \ell-3)}{4} \end{equation}

 When we comparing Eq.(\ref{eq:gc9}) and Eq.(\ref{eq:gc1}) we get

\begin{equation} \tilde{\tau}(z)=2z ,\ \ \ \   \sigma(z)=z^2 \label{eq:gc10}\end{equation}
\begin{equation} \tilde{\sigma}(z)=-\varepsilon+\eta z-\gamma z^2 \label{eq:gc11}\end{equation}
\begin{equation} \sigma^{'}(z)=2 z, \ \ \ \  \sigma^{''}(z)=2 \label{eq:gc12}\end{equation}
We substitute Eq. (\ref{eq:gc10}), (\ref{eq:gc11}),(\ref{eq:gc12}) and Eq.(\ref{eq:gc2}) and obtain
\begin{equation} \pi(z)=\pm \sqrt{\varepsilon-\eta z +(\gamma+k) z^2} \label{eq:gc13} \end{equation}

 To find $k$, we take the function's discriminant under the square root, which gives us

\begin{equation} k=\frac{\eta^2-4 \gamma \varepsilon}{4 \varepsilon} \label{eq:gc14} \end{equation}

 We substitute Eq.(\ref{eq:gc14}) into Eq.(\ref{eq:gc13}) and have

\begin{equation} \pi(z)=\pm \frac{1}{\sqrt{\varepsilon}}\left(\frac{nz}{2}-\varepsilon\right)\label{eq:gc15}  \end{equation}

 We differentiate the negative component of Eq.(\ref{eq:gc15}) which gives us
\begin{equation} \pi_{-}^{'}(z)=-\frac{\eta}{2 \sqrt{\varepsilon}} \label{eq:gc16}  \end{equation}

 By substituting Eq.(\ref{eq:gc10}), (\ref{eq:gc11}),(\ref{eq:gc12}) and (\ref{eq:gc16}  ) into Eq.(\ref{eq:gc1b}) We've got

 \begin{equation} \tau(z)= 2z-\frac{\eta z}{\sqrt{\varepsilon}}+\frac{2 \varepsilon}{\sqrt{\varepsilon}} \label{eq:gc17}  \end{equation}
 Differentiating Eq.(\ref{eq:gc17}) We've got

 \begin{equation} \tau^{'}(z)=2-\frac{\eta}{ \sqrt{\varepsilon}} \end{equation}

With the help of Eq.(\ref{eq:gc2c}) we obtain

\begin{equation}  \lambda=\frac{\eta^2-4 \gamma \varepsilon}{4 \varepsilon}-\frac{\eta}{2 \sqrt{\varepsilon}}\label{eq:gc18} \end{equation}
  Using Eq. (\ref{eq:gc2d}), we obtain

\begin{equation} \lambda_{n}=\frac{n \eta}{\sqrt{\varepsilon}}-n^2-n \label{eq:gc19} \end{equation}

 Equating Eq.(\ref{eq:gc18}) and (\ref{eq:gc19}) substituting Eq.(\ref{eq:gc3a}) and Eq.(\ref{eq:gc9}) gives the LIMYP energy eigenvalue equation in the relativistic limit as, Considering a transformation of the form: $M+E_{n \ell}\rightarrow \frac{2 \mu}{\hbar^2}$ and $ M-E_{n \ell} \rightarrow -E_{n \ell} $ where $\mu$ is reduced mass. The non-relativistic energy eigenvalues equation is as follows,

\[ E_{n \ell}=-\frac{(6 A_{1}+A_{2} \alpha^2+4 A_{3} \alpha^3)}{\delta}-\frac{A_{2} \alpha^3}{\delta^2}-2(A_{4}-2 A_{3}\alpha^3- A_{2} \alpha)-\frac{\hbar^2}{8 \mu}\]
\begin{equation}\left[\frac{\frac{4\mu}{\hbar^2}(A_{2}+2 A_{3} \alpha)+\frac{2 \mu}{\hbar^2 \delta^2}(6A_{1}+A_{2}\alpha^2+4 A_{3}\alpha^3)-\frac{16 \mu A_{2} \alpha^3}{3 \hbar^2 \delta^3}}{n+\frac{1}{2}+\sqrt{\frac{1}{4}+\frac{4 A_{3}\mu}{\hbar^2}+\frac{4\mu}{\hbar^2 \delta^3}(A_{1}+\frac{A_{2}\alpha^2}{2}+\frac{4}{3}A_{3} \alpha^3)+\frac{A_2 \alpha^3 \mu}{\hbar^2 \delta^4}+\frac{(N+2 \ell-1)(N+2 \ell-3)}{4}}}\right]^2 \label{eq:gc.19}\end{equation}

 To find out the wavefunction, we substitute Eqs.(\ref{eq:gc10}),(\ref{eq:gc11}) (\ref{eq:gc12}) and (\ref{eq:gc15}) into Eq.(\ref{eq:gc1d}) and obtain

 \begin{equation} \frac{d\phi}{\phi}=\left(\frac{\varepsilon}{z^2\sqrt{\varepsilon}}-\frac{\alpha}{2z\sqrt{\varepsilon}}\right)dx \label{eq:gc20} \end{equation}

We integrating Eq.(\ref{eq:gc20}) and we get
\begin{equation} \phi(z)= z^{-\frac{\alpha}{2\sqrt{\varepsilon}}} e^{-\frac{\varepsilon}{z\sqrt{\varepsilon}}} \label{eq:gc21} \end{equation}

By substituting Eqs.(\ref{eq:gc10}),(\ref{eq:gc11}) (\ref{eq:gc12}) and (\ref{eq:gc1d}) into Eq.(\ref{eq:gc1f}) we integrating and then simplify we obtain

\begin{equation} \rho(z)= z^{-\frac{\alpha}{\sqrt{\varepsilon}}} e^{-\frac{2 \varepsilon}{z\sqrt{\varepsilon}}} \label{eq:gc21aa} \end{equation}

 We Substituting Eq.(\ref{eq:gc10}), (\ref{eq:gc11}), (\ref{eq:gc12}) and (\ref{eq:gc21aa}) into Eq.(\ref{eq:gc1e}) We've got

\begin{equation} y_{n}(z)=B_{n}e^{\frac{2\varepsilon}{z\sqrt{\varepsilon}}}z^{\frac{\alpha}{\sqrt{\varepsilon}}}\frac{d^{n}}{dz^{n}}\left[e^{-\frac{2\varepsilon}{z\sqrt{\varepsilon}}}z^{2n-\frac{\alpha}{\sqrt{\varepsilon}}}\right]\end{equation}

The associated Laguerre polynomials have the Rodrigue's formula.
\begin{equation} L_{n}^{\frac{\alpha}{\sqrt{\varepsilon}}}\left[\frac{2\varepsilon}{z\sqrt{\varepsilon}}\right]=\frac{1}{n!}e^{\frac{2\varepsilon}{z\sqrt{\varepsilon}}}z^{{\frac{\alpha}{\sqrt{\varepsilon}}}}\frac{d^n}{dz^n}\left(e^{-\frac{2\varepsilon}{z\sqrt{\varepsilon}}}z^{2n-\frac{\alpha}{\sqrt{\varepsilon}}}\right)\end{equation}

Where

\begin{equation} \frac{1}{n!}=B_{n}  \label{eq:gc22} \end{equation}

 Hence,

\begin{equation}y_{n}(z)\equiv L_{n}^{\frac{\alpha}{\sqrt{\varepsilon}}}\left(\frac{2\varepsilon}{z\sqrt{\varepsilon}}\right)\end{equation}

We substituting Eqs.(\ref{eq:gc21}) and (\ref{eq:gc22}) into Eq.(\ref{eq:gc1h}) we get the wavefunction of Eq.(\ref{eq:gc6}) in terms of Lagurre polynomial as

\begin{equation} \psi(z)=B_{nl}z^{-\frac{\alpha}{2\sqrt{\varepsilon}}}e^{-\frac{\varepsilon}{z\sqrt{\varepsilon}}}L_{n}^{\frac{\alpha}{\sqrt{\varepsilon}}}\left(\frac{2\varepsilon}{z\sqrt{\varepsilon}}\right)\end{equation}

Where $ N_{nl}$ is normalization constant, which may be determined by

\begin{equation}\int_{0}^{\infty}|B_{nl}(r)|^2 dr=1 \end{equation}

\section{Mass spectra}
We modify the mass spectra of heavy mesons with quark and antiquark flavors, such as charmonium, Bottomonium, and $b\bar{c}$. The following equation is used to calculate the mass spectra.

\begin{equation} M=m_{1}+m_{2}+E_{n \ell} \end{equation}
but,
\begin{equation} m_{1}=m_{2}=m_{b} \end{equation}

As a result of this, the expression

\begin{equation} M=2m_{b}+E_{n \ell} \label{eq:gc.22a} \end{equation}
where $m_b$ is the mass of the particle being studied and $ E_{n\ell}$ is the calculated energy eigenvalues
 We substituting Eq.(\ref{eq:gc.22a}) into Eq.(\ref{eq:gc.19}) we obtain
\[m=2m+ \left(-\frac{(6 A_{1}+A_{2} \alpha^2+4 A_{3} \alpha^3)}{\delta}\right)-\frac{A_{2} \alpha^3}{\delta^2}-2(A_{4}-2 A_{3}\alpha^3- A_{2} \alpha)-\frac{\hbar^2}{8 \mu}\]
\begin{equation}\left[\frac{\frac{4\mu}{\hbar^2}(A_{2}+2 A_{3} \alpha)+\frac{2 \mu}{\hbar^2 \delta^2}(6A_{1}+A_{2}\alpha^2+4 A_{3}\alpha^3)-\frac{16 \mu A_{2} \alpha^3}{3 \hbar^2 \delta^3}}{n+\frac{1}{2}+\sqrt{\frac{1}{4}+\frac{4 A_{3}\mu}{\hbar^2}+\frac{4\mu}{\hbar^2 \delta^3}(A_{1}+\frac{A_{2}\alpha^2}{2}+\frac{4}{3}A_{3} \alpha^3)+\frac{A_2 \alpha^3 \mu}{\hbar^2 \delta^4}+\frac{(N+2 \ell-1)(N+2 \ell-3)}{4}}}\right]^2 \label{eq:gc23}\end{equation}



 { \begin{table}

  \caption{\label{math-tab2}The mass spectra of  $c{\bar c}$ ($M_{c{\bar c}}$) in GeV  for $m_c=1.209$ GeV, $\hbar=1$, $N=3$,   $A_1=0.100685 $ GeV, $A_2=0.233222$ GeV , $A_3=0.0010124$ GeV, $A_{4}=0.001012$ GeV }

\begin{tabular*}{\textwidth}{@{}l*{15}{@{\extracolsep{0pt plus
12pt}}l}}
\br
  State  &       Our work  &  \cite{gc.22} & \cite{soni} & \cite{gc.25}
   &  \cite{gc.29b}    & Exp.  \cite{gc.30}   \\
   \br

  1S &  3.096 & 3.096 & 3.094& 2.950 & 3.096 & 3.096\\

 2S & 3.733 & 3.686 &3.681 & 3.522 & 3.672 & 3.686\\

 1P &  3.415  &  3.255 & 3.681& 3.398 & 3.521& 3.415\\

2P & 3.894 & 3.779 &3.870 &  3.792 & 3.951 & 3.918 \\

3S & 4.068 & 4.040 &4.058 & 3.912 & 4.085 & 4.085  \\

4S & 4.263  &4.269 & 4.448 &-& 4.433 &4.263 \\

 1D&3.770 &3.504  &3.772 &-&3.779&3.770\\

 2D & 4.088&- &4.188&  - & 4.159 & - \\
 1F& 4.040&-& 4.017& & -& -\\

  \br

\end{tabular*}

\end{table}}

\begin{figure}[!tbp]
  \centering
  \begin{minipage}[b]{0.45\textwidth}
  \includegraphics[width=1.1\textwidth]{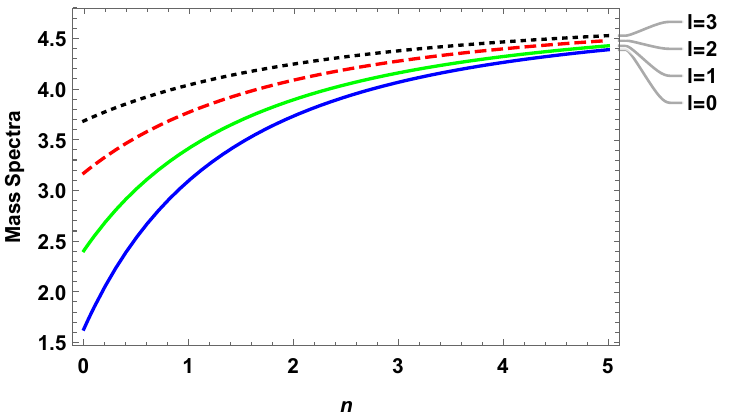}
  \caption{Graph of the   $M_{c{\bar c}}$ $\rightarrow$ principal quantum number ($n$).}\label{fig:ch.1}
 \end{minipage}
  \hfill
  \begin{minipage}[b]{0.45\textwidth}
  \includegraphics[width=1.1\textwidth]{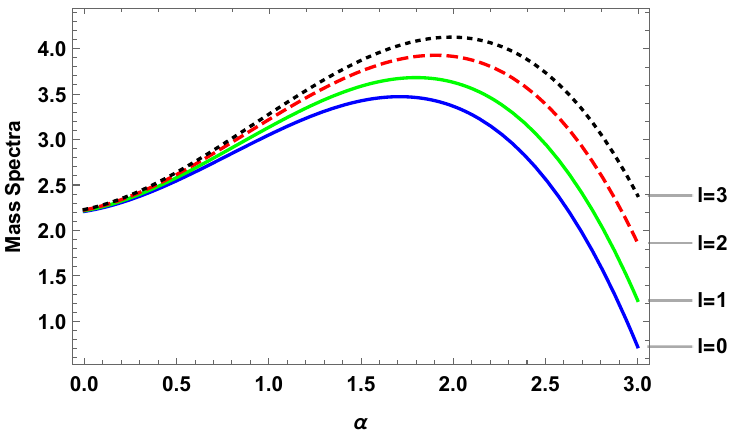}
  \caption{Graph of the $M_{c{\bar c}}$  $\rightarrow$ screening parameter ($\alpha$)  with different azimuthal quantum number ($\ell$).}\label{fig:ch.2}
 \end{minipage}
\end{figure}

\begin{figure}[!tbp]
  \centering
  \begin{minipage}[b]{0.45\textwidth}
  \includegraphics[width=1.1\textwidth]{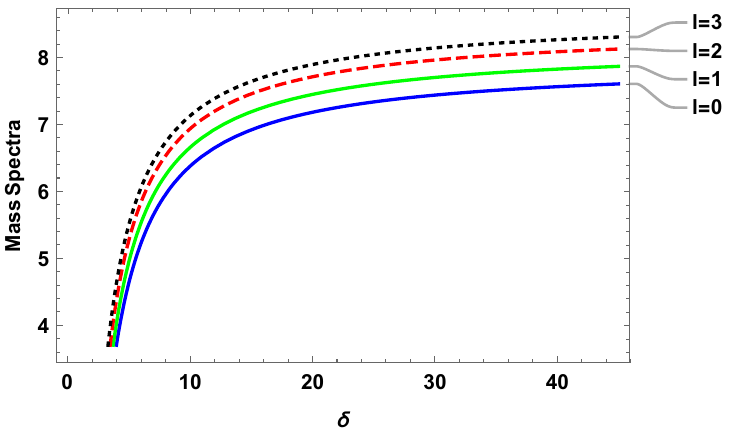}
  \caption{ Graph of the   $M_{c{\bar c}}$   $\rightarrow$ $\delta$ with different azimuthal quantum number ($\ell$).}\label{fig:ch.3}
 \end{minipage}
  \hfill
  \begin{minipage}[b]{0.45\textwidth}
  \includegraphics[width=1.1\textwidth]{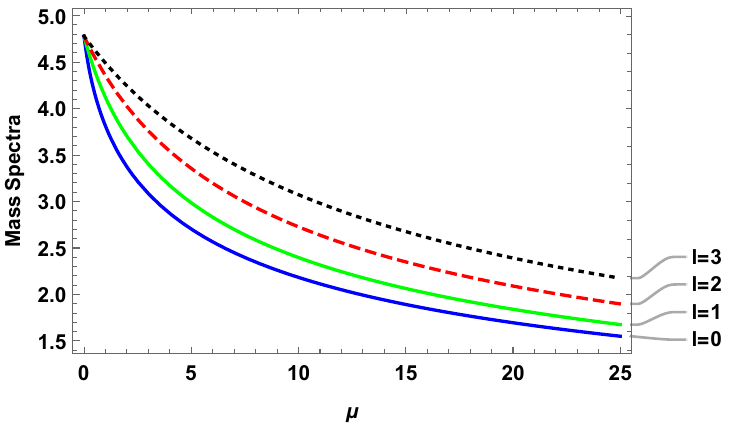}
  \caption{Graph of the  $M_{c{\bar c}}$   $\rightarrow$ reduced mass ($\mu$) with different azimuthal quantum number ($\ell$).}\label{fig:ch.4}
 \end{minipage}
\end{figure}

\begin{figure}[!tbp]
  \centering
  \begin{minipage}[b]{0.45\textwidth}
  \includegraphics[width=1.1\textwidth]{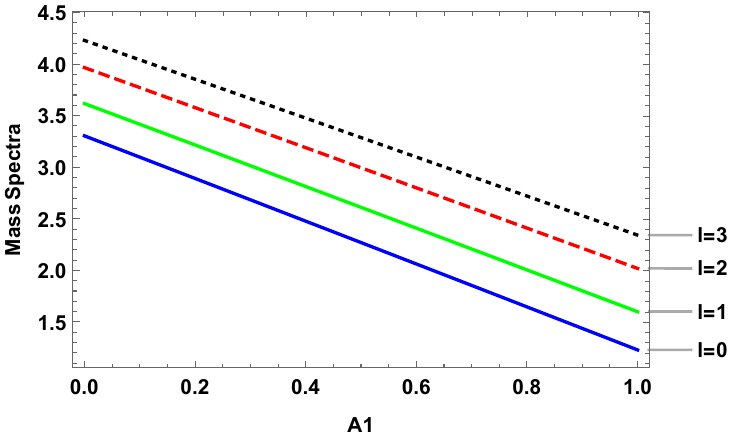}
  \caption{Graph of the $M_{c{\bar c}}$ $\rightarrow$ potential strength ($A_1$) with different azimuthal quantum number ($\ell$).}\label{fig:ch.5}
 \end{minipage}
  \hfill
  \begin{minipage}[b]{0.45\textwidth}
  \includegraphics[width=1.1\textwidth]{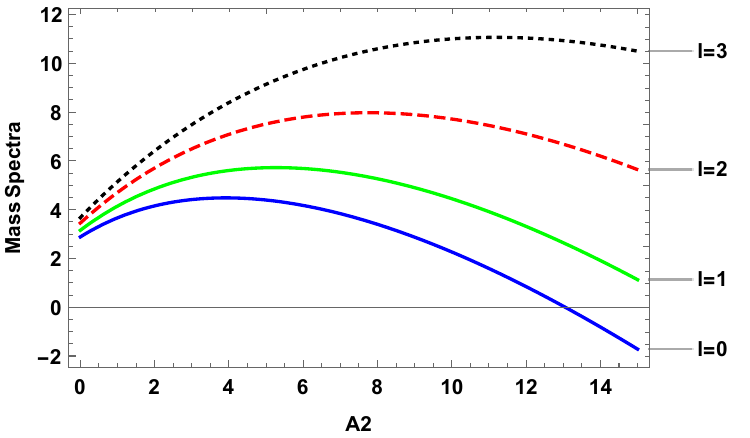}
  \caption{Graph of the  $M_{c{\bar c}}$ $\rightarrow$  potential strength ($A_2$) with different azimuthal quantum number ($\ell$).}\label{fig:ch.6}
 \end{minipage}
\end{figure}

\begin{figure}[!tbp]
  \centering
  \begin{minipage}[b]{0.45\textwidth}
  \includegraphics[width=1.1\textwidth]{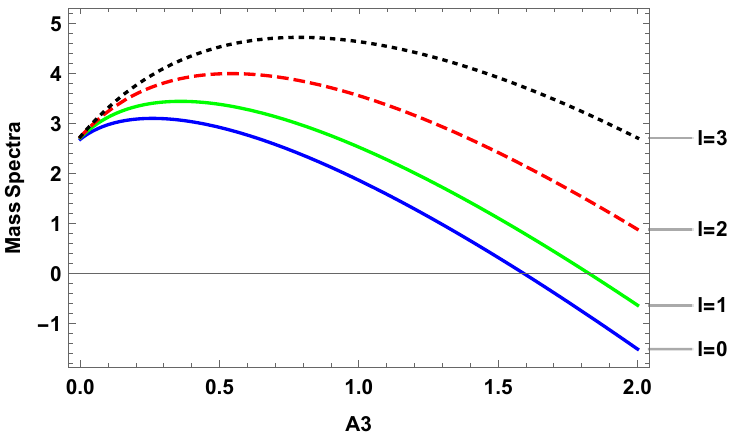}
  \caption{Graph of the  $M_{c{\bar c}}$  $\rightarrow$ potential strength ($A_3$) with different azimuthal quantum number ($\ell$).}\label{fig:ch.7}
 \end{minipage}
  \hfill
  \begin{minipage}[b]{0.45\textwidth}
  \includegraphics[width=1.1\textwidth]{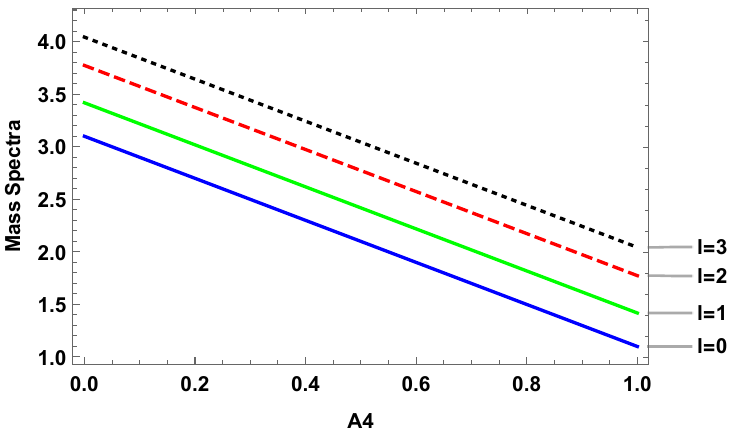}
  \caption{Graph of the $M_{c{\bar c}}$  $\rightarrow$ potential strength ($A_4$) with  different azimuthal quantum number ($\ell$).}\label{fig:ch.8}
 \end{minipage}
\end{figure}

{\begin{table}
  \caption{\label{math-tab2}The mass spectra of   $b{\bar b}$ $(M_{b{\bar b}})$ in GeV  for $m_b=4.823$, $\hbar=1$, $N=3$,   $A_1=0.100685$ GeV, $A_2=0.20752$ GeV , $A_3=0.270076$ GeV, $A_{4}=0.0010124$ GeV }
\begin{tabular*}{\textwidth}{@{}l*{15}{@{\extracolsep{0pt plus
12pt}}l}}
\br
  State  &       Our work  &  \cite{gc.22} &  \cite{soni} & \cite{gc.25}
   &  \cite{gc.29b}    & Exp. \cite{gc.30}   \\
   \br

  1S &  9.460 & 9.460&9.463 & 9.411 & 9.462 & 9.460\\

 2S & 10.028 & 10.023&9.979 & 9.826 & 10.027 & 10.023\\

  1P &  9.704 & 9.619& 9.821 & 9.755 & 9.963 & 9.899\\

2P & 10.160 & 10.114 &10.220& 10.035 & 10.299 & 10.260 \\

3S &10.343& 10.355&10.359 & 10.088 & 10.361 & 10.355 \\

4S & 10.536 &10.567 &10.683& -& 10.624 &10.580 \\

 1D&10.010&9.864&10.074&-&10.209&10.164\\

 2D & 10.332 &  -&10.441&- & - &-& \\

1F & 10.268 & -&10.288 & - & - & \\

  \br

\end{tabular*}
\end{table}}

\begin{figure}[!tbp]
  \centering
  \begin{minipage}[b]{0.45\textwidth}
  \includegraphics[width=1.1\textwidth]{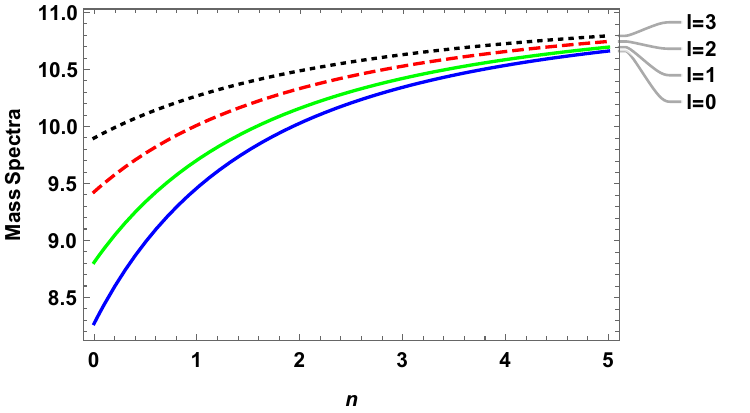}
  \caption{ Graph of the  $M_{b{\bar b}}$ $\rightarrow$ principal quantum number ($n$).}\label{fig:bt.1}
 \end{minipage}
  \hfill
  \begin{minipage}[b]{0.45\textwidth}
  \includegraphics[width=1.1\textwidth]{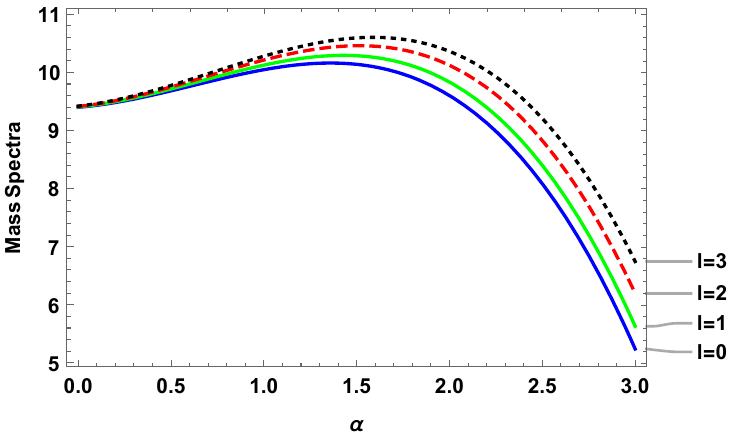}
  \caption{Graph of the  $M_{b{\bar b}}$  $\rightarrow$ screening parameter $\alpha$ with different azimuthal quantum number ($\ell$).}\label{fig:bt.2}
 \end{minipage}

\end{figure}

\begin{figure}[!tbp]
  \centering
  \begin{minipage}[b]{0.45\textwidth}
  \includegraphics[width=1.1\textwidth]{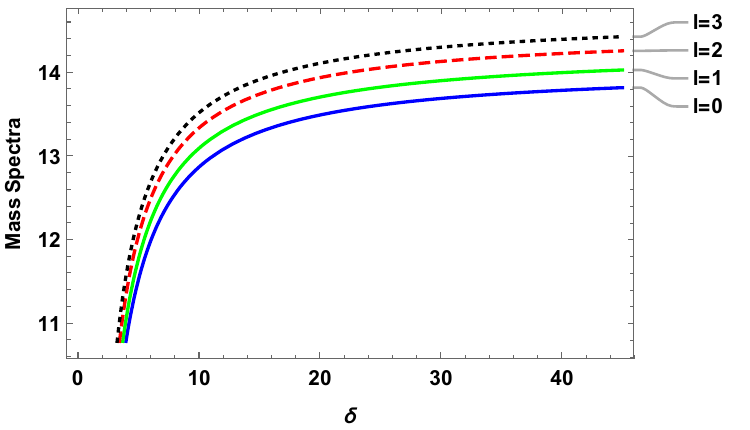}
  \caption{ Graph of the $M_{b{\bar b}}$  $\rightarrow$ $\delta$ with different azimuthal quantum number ($\ell$).}\label{fig:bt.3}
 \end{minipage}
  \hfill
  \begin{minipage}[b]{0.45\textwidth}
  \includegraphics[width=1.1\textwidth]{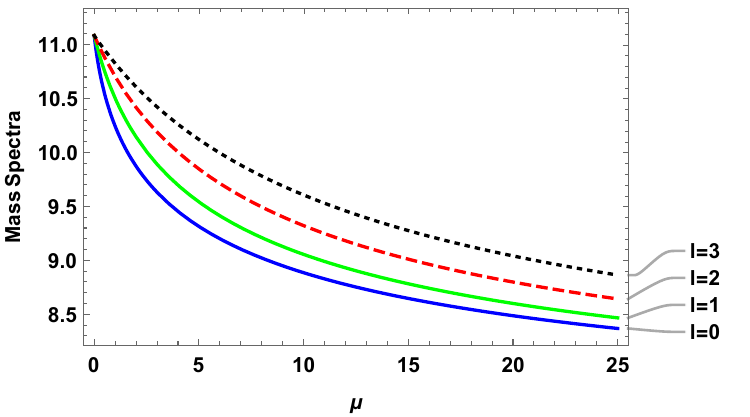}
  \caption{Graph of the  $M_{b{\bar b}}$  $\rightarrow$ reduced mass ($\mu$) with different azimuthal quantum number ($\ell$).}\label{fig:bt.4}
 \end{minipage}
\end{figure}

\begin{figure}[!tbp]
  \centering
  \begin{minipage}[b]{0.45\textwidth}
  \includegraphics[width=1.1\textwidth]{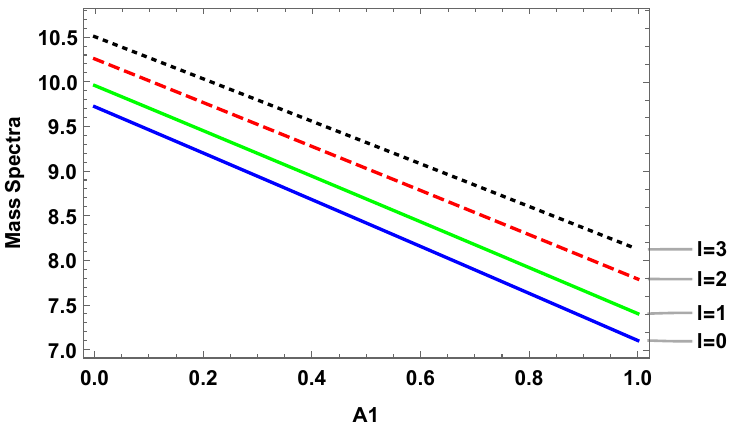}
  \caption{ Graph of the  $M_{b{\bar b}}$  $\rightarrow$ potential strength ($A_1$) with different azimuthal quantum number ($\ell$).}\label{fig:bt.5}
 \end{minipage}
  \hfill
  \begin{minipage}[b]{0.45\textwidth}
  \includegraphics[width=1.1\textwidth]{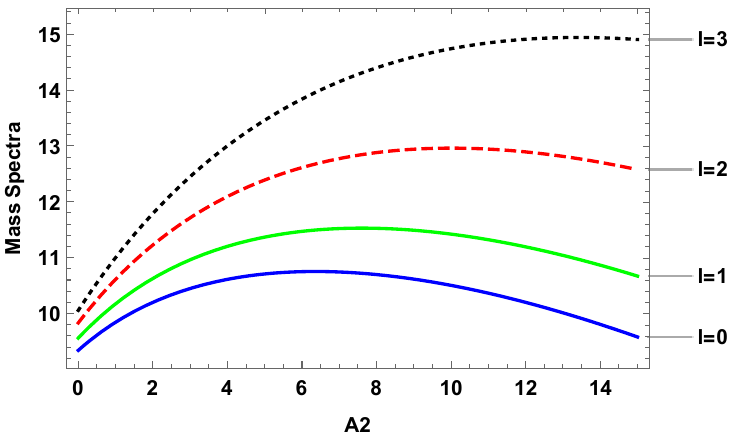}
  \caption{Graph of the $M_{b{\bar b}}$ $\rightarrow$ potential strength ($A_2$) with different azimuthal quantum number ($\ell$).}\label{fig:bt.6}
 \end{minipage}
\end{figure}

\begin{figure}[!tbp]
  \centering
  \begin{minipage}[b]{0.45\textwidth}
  \includegraphics[width=1.1\textwidth]{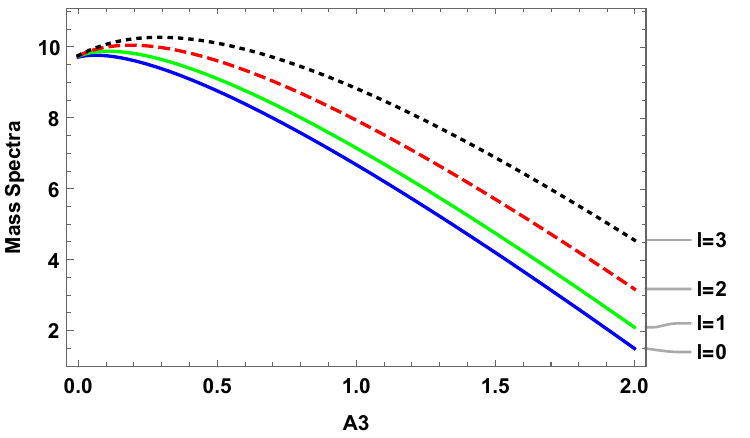}
  \caption{Graph of the  $M_{b{\bar b}}$  $\rightarrow$ potential strength($A_3$) with different azimuthal quantum number ($\ell$).}\label{fig:bt.7}
 \end{minipage}
  \hfill
  \begin{minipage}[b]{0.45\textwidth}
  \includegraphics[width=1.1\textwidth]{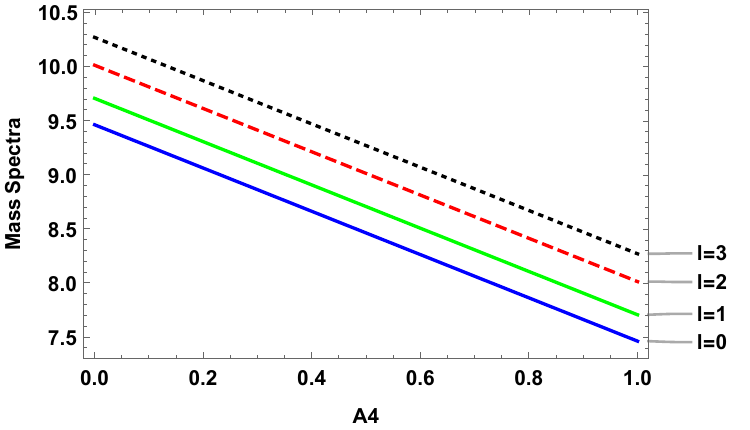}
  \caption{Graph of the $M_{b{\bar b}}$  $\rightarrow$  potential strength ($A_4$) with different azimuthal quantum number ($\ell$).}\label{fig:bt.8}
 \end{minipage}
\end{figure}

 { \begin{table}

  \caption{\label{math-tab2} The mass spectra of   $b\bar{c}$ $(M_{b\bar{c}})$   for parameters $m_c=1.209$  GeV, $m_{b}=4.823$  GeV, $\hbar=1$, $N=3$, $A_1=0.400685$ GeV, $A_2=0.070752$ GeV, $A_3=0.0705700$ GeV, $A_{4}=0.4101241$  GeV }

\begin{tabular*}{\textwidth}{@{}l*{15}{@{\extracolsep{0pt plus
12pt}}l}}
\br
  State  &       Our work & \cite{gc.26} &  \cite{gc.31}
   &  \cite{gc.31a} & \cite{gc.31b}  & Exp.\cite{gc.32}     \\
\br

  1S &  6.277& 6.318 & 6.277 & 6.349 & 6.270 & 6.275\\

 1P & 6.604 &6.769 & 6.340  & 6.715 & 6.699&\\

2S & 6.961& 6.870 & 6.814  & 6.821 & 6.835&  6.842 \\

2P& 7.130 & 7.175 &6.851 & 7.102 & 7.091 & - \\

3S & 7.326& 7.248 & 7.351 & 7.175 & 7.193& \\
4S & 7.543 &7.567 & 7.889 & - & -& \\

1D & 6.984&7.028 & - & -& -& \\

2D & 7.339 & 7.365 & - & - & - & \\

1F& 7.281 & - & -& - &\\

 \br

\end{tabular*}

\end{table}}

\begin{figure}[!tbp]
  \centering
  \begin{minipage}[b]{0.45\textwidth}
  \includegraphics[width=1.1\textwidth]{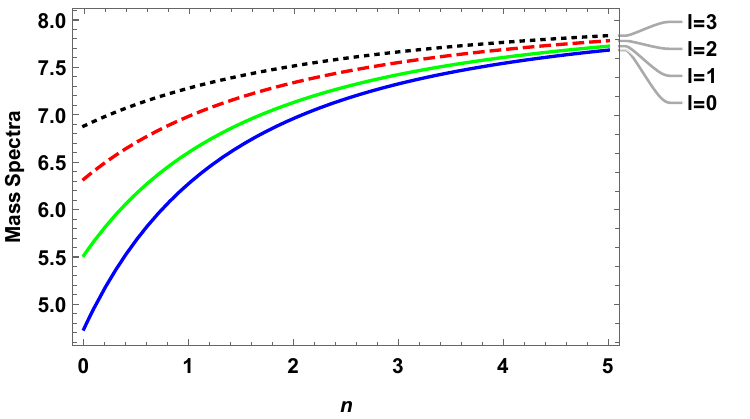}
  \caption{Graph of the $M_{b\bar{c}}$  $\rightarrow$principal quantum number ($n$) .}\label{fig:bc.1}
 \end{minipage}
  \hfill
  \begin{minipage}[b]{0.45\textwidth}
  \includegraphics[width=1.1\textwidth]{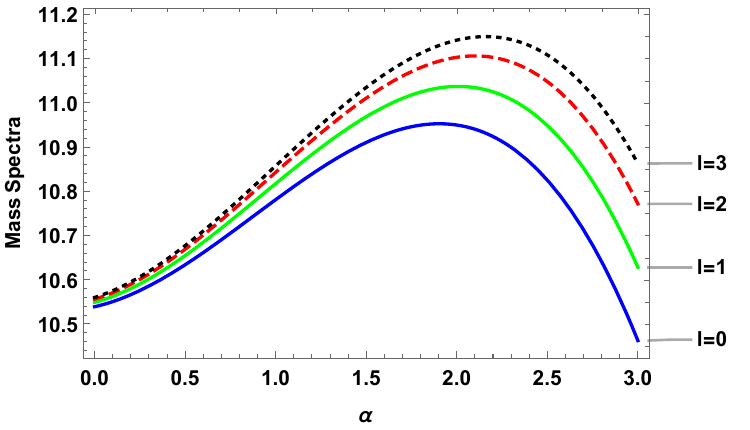}
  \caption{Graph of the $M_{b\bar{c}}$  $\rightarrow$ screening parameter ($\alpha$) with different azimuthal quantum number ($\ell$).}\label{fig:bc.2}
 \end{minipage}
\end{figure}

\begin{figure}[!tbp]
  \centering
  \begin{minipage}[b]{0.45\textwidth}
  \includegraphics[width=1.1\textwidth]{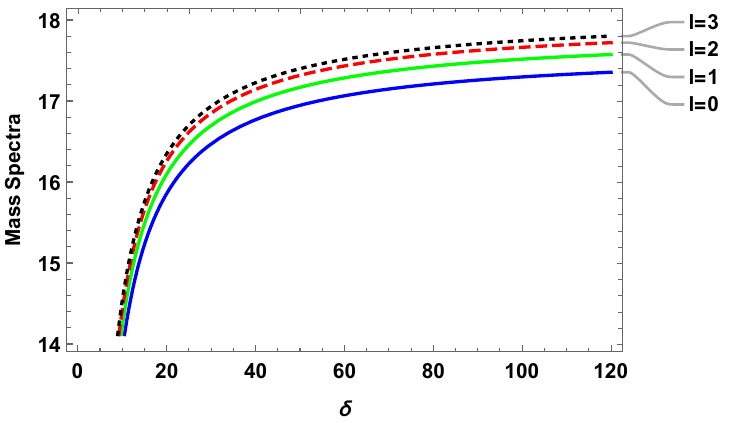}
  \caption{ Graph of the  $M_{b\bar{c}}$  $\rightarrow$ $\delta$ with different azimuthal quantum number ($\ell$).}\label{fig:bc.3}
 \end{minipage}
  \hfill
  \begin{minipage}[b]{0.45\textwidth}
  \includegraphics[width=1.1\textwidth]{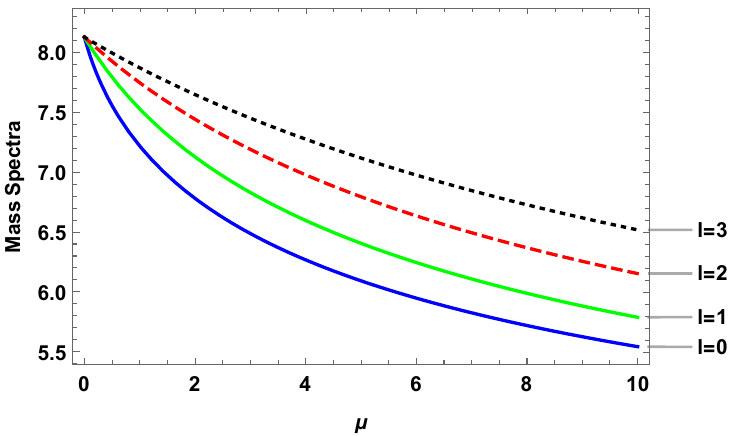}
  \caption{ Graph of the $M_{b\bar{c}}$  $\rightarrow$ reduced mass ($\mu$) with different azimuthal quantum number ($\ell$).}\label{fig:bc.4}
 \end{minipage}
\end{figure}

\begin{figure}[!tbp]
  \centering
  \begin{minipage}[b]{0.45\textwidth}
  \includegraphics[width=1.1\textwidth]{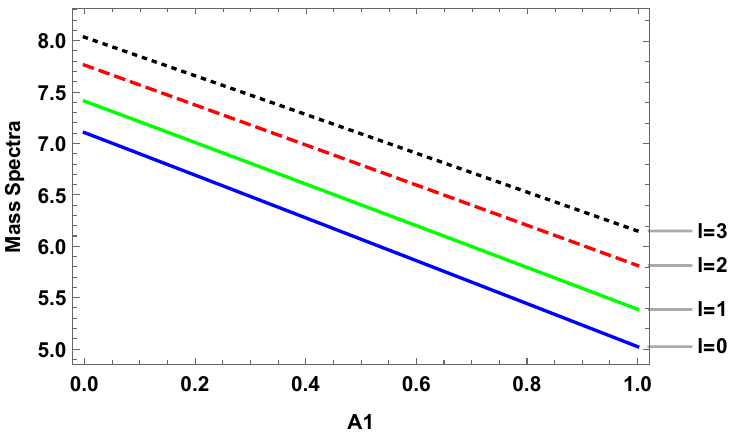}
  \caption{ Graph of the  $M_{b\bar{c}}$  $\rightarrow$ potential strength ($A_1$) with different azimuthal quantum number ($\ell$).}\label{fig:bc.5}
 \end{minipage}
  \hfill
  \begin{minipage}[b]{0.45\textwidth}
  \includegraphics[width=1.1\textwidth]{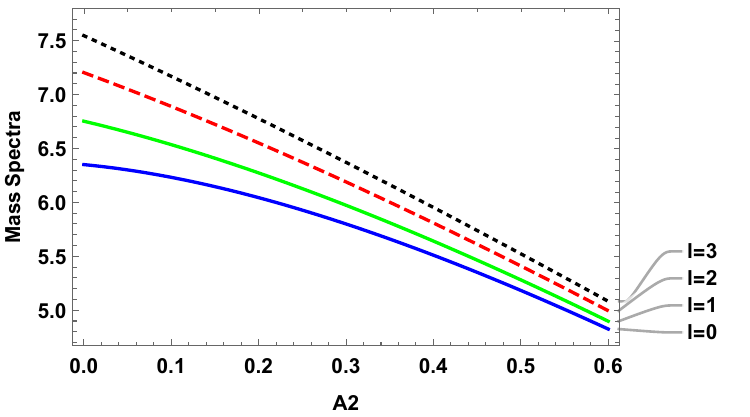}
  \caption{Graph of the $M_{b\bar{c}}$  $\rightarrow$ reduced with potential strength ($A_2$) with  different azimuthal quantum number ($\ell$).}\label{fig:bc.6}
 \end{minipage}
\end{figure}

\begin{figure}[!tbp]
  \centering
  \begin{minipage}[b]{0.45\textwidth}
  \includegraphics[width=1.1\textwidth]{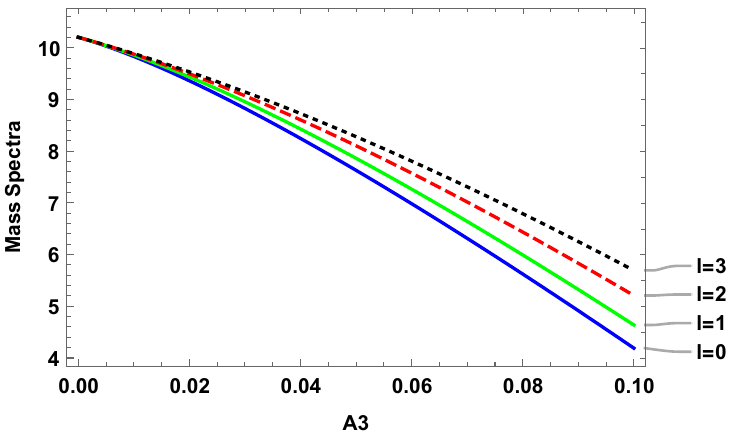}
  \caption{Graph of the  $M_{b\bar{c}}$ \unboldmath $\rightarrow$ potential strength ($A_3$) with different azimuthal quantum number ($\ell$).}\label{fig:bc.7}
 \end{minipage}
  \hfill
  \begin{minipage}[b]{0.45\textwidth}
  \includegraphics[width=1.1\textwidth]{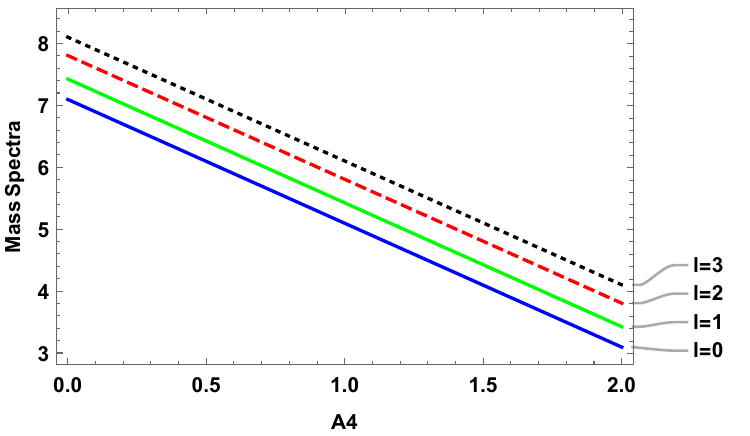}
  \caption{Graph of the $M_{b\bar{c}}$ $\rightarrow$ potential strength ($A_4$) with different azimuthal quantum number ($\ell$).}\label{fig:bc.8}
 \end{minipage}
\end{figure}

\section{Discussion of Results}
The bound state solution of the K-G equation has been successfully used to determine the mass spectra of $c\bar{c}$, $b\bar{b}$, $b\bar{c}$ for states ranging from $1S$, $2S$, $1P$, $2P$, $3S$, $4S$, $1D$, $2D$, and $1F$ by using Eq.(\ref{eq:gc23}) which includes linear plus modified Yukawa potential, which finds good co-relation of mass spectra with the experimental data and \cite{gc.30,gc.32} as well as some recent research \cite{gc.22,gc.31a}.  Our results for charmonium mass spectra are shown in Table. 1: The obtained results are compared to PDG \cite{gc.30}   results as well as other theoretical findings. The obtained mass spectra of $1S, 1P, 1D, and 4S$ states are in very good agreement with PDG\cite{gc.30} observation. We observed that the experimental value for the 2S state which has a mass of 3.686 GeV is reasonably close to the predicted mass of 3.733 GeV with a mass difference of 47 MeV.  Experimentally, the two-state 2P and 3S masses are 3.918 and 4.085 GeV, respectively, which are very close to our calculated masses of 3.894 and 4.067 with a mass difference of 24 and 18 MeV respectively. In Table. 2: the mass spectra of $b{\bar{b}}$ system is shown along with PDG \cite{gc.30} as well as with other theoretical predictions. Experimentally found 2S, 3S, and 4S states with masses of 10.023, 10.355, and 10.164 GeV are near to our calculated masses of 10.028, 10.343, and 10.535, respectively, with mass differences of 5, 12, and 45 MeV.  In Table. 3:  we observed that the experimentally found state 1S having a mass of 6.275 GeV is also close to our predicted mass of 6.277 GeV with a slight mass difference of 2 MeV.  Various plots of mass spectra with respect to all the variables used in the energy equation (\ref{eq:gc23}) such as principal quantum number $(n)$, reduce mass $(\mu)$, screening parameter $(\alpha)$,  $\delta$  and potential strength $A_{1}, A_{2}, A_{3}, A_{4}$.  From Fig.(1), (9), and (17) we depicted that for any sub-shell $S, P, D, F$ all show a similar pattern of increasing mass spectra with an increasing principal quantum number. Which in agreement with experimental data used in Table.1, 2, and 3  for different sub-shells with increasing principal quantum number mass spectra increases respectively. We show in Figures (2), (10), and (18) that for any constant sub-shell $S, P, D, F$, all show a similar trend of increasing up to a certain screening parameter value and then declining, with any further increase in screening parameter depending on quark anti-quark interaction potential.  In Fig. (3), (11), and (19) we show that for any constant sub-shell $S, P, D, F$ all show a similar pattern of first abruptly increasing mass spectra followed by the continuous increase in mass spectra with an increase $\delta$.  In case of reduced mass Fig. (4), (12), and (20)  for any constant sub-shell $S, P, D, F$ all show a similar exponentially decreasing pattern with an increase in reduced mass. For all the heavy mesons from Fig.(5), (8), (13),  (16), (21), and (24) we observed that for any constant sub-shell $S, P, D, F$ potential strength $A_{1}$ and $A_{4}$ decreases linearly. In case of potential strength, $A_{2}$ and $A_{3}$  for charmonium and bottomonium fig.(6), (7), (14) and (15) there are slightly different in the graph of mass spectra as compared to its counter potential strength $A_{1}$ and $A_{4}$ initially it shows a  hump of increase in mass spectra for a certain potential which eventually decreases but for ${b \bar{c}}$ there is no initial increase they decay similarly as $A_{1}$ and $A_{4}$ while considering the case of $A_{2}$ for constant sub-shell $S, P, D, F$ initially there is a large difference in mass spectra which on increasing on potential strength $A_{2}$ becomes similar for all sub-shell. In the case of $A_{3}$ initially all the constant sub shell  $S, P, D, F$ have similar mass spectra which on increasing potential strength $A_{3}$ as sub-shell decreases with a high slope as compared to $P, D, F$ sub-shell, resulting in a gap in mass spectra of all sub-shell on higher potential strength $A_{3}$. The figure shown in this paper will help to understand the behavior of the linear plus modified  Yukawa potential.

\section{Conclusion}
 Here, the linear plus modified Yukawa potential has been successfully used to calculate the mass spectra of heavy-heavy flavored mesons. The mass spectra of $c\bar{c}$, $b\bar{b}$, and $b\bar{c}$ are determined using the energy eigenvalue expression and listed in table 1, 2, and 3, respectively wherein we have attempted to compare the present results with other theoretical findings as well as experimental states (wherever available). Different plots of mass spectra versus different potential factors were drawn as depicted in section 3. In the energy equation for all heavy mesons, the variation of mass spectra with the following parameter is employed, such that mass spectra increase with an increasing principal quantum number in Fig.(1), Fig.(9), and Fig.(17).  In Fig.(2), Fig.(10), and Fig.(18), mass spectra first increase and then decrease as the screening parameter is increased.  Mass spectra grow as delta increases in In Fig.(3), Fig.(11), and Fig.(19). The mass spectra in Fig.(4), Fig.(12), and Fig.(20) exponentially decrease with increased reduce mass. With rising potential strength $A_{1}$ and $A_{4}$ correspondingly, mass spectra drop linearly. While the variation of mass spectra with potential strength $A_2$ and $A_3$ first increases and drops for $c\bar{c}$ and $b\bar{b}$, respectively. Furthermore, the trend of mass spectra with potential strengths $A_2$ and $A_3$ decreases for $b\bar{c}$. The overall spectra of $c\bar{c}$, $b\bar{b}$, and $b\bar{c}$ mesons are in good agreement with experimental data and previous theoretical investigations. As linear plus modified Yukawa potential  Eq.(\ref{eq:gc30}) incorporates the two screening terms (+ve and -ve) is expected to suppress the mass spectra of higher excited states of those mesons $(c{\bar{c}}, b{\bar{b}}, b{\bar{c}})$. We can see this effect in $b{\bar{b}}$ and $c{\bar{b}}$ meson where the higher state mass spectra are suppressed due to screening effect with few MeV with experimental as well as other theoretical results. We would like to extend this model to calculate the mass spectra of heavy-light flavored mesons. The potential utilized in this study can also be reduced in order to derive an energy equation for diatomic molecules.\\ \\


\textbf{Acknowledgment}\\

The authors are thankful to the ICNFP 2021 organizing committee for allowing us to present our work.

\end{document}